\documentclass[12pt]{elsarticle}

\pdfoutput=1

\let\eqref\relax

\usepackage{amssymb}
\usepackage{amsthm}

\usepackage{pstricks}
\usepackage{pst-node}
\usepackage{url}
\usepackage{subfig}

\usepackage{dcpic,pictexwd}

\usepackage{courier}

\usepackage{lineno}
\usepackage{url}
\usepackage{graphicx}
\graphicspath{ {./Images/} }

\usepackage{amsmath}
\DeclareMathOperator*{\argmin}{arg\,min}
\journal{Computers and Mathematics with Applications}

\let\oldequation\equation
\let\oldendequation\endequation

\renewenvironment{equation}
  {\linenomathNonumbers\oldequation}
  {\oldendequation\endlinenomath}

\begin{document}

\begin{frontmatter}



\title{Cuts for 3-D magnetic scalar potentials: visualizing unintuitive surfaces arising from trivial knots}


\author[BU]{Alex Stockrahm}
\ead{adstockrahm@gmail.com}
\ead[url]{http://bu.edu}
\author[TUNI]{Valtteri Lahtinen}
\ead{valtteri.lahtinen@tuni.fi}
\ead[url]{http://www.tuni.fi}
\author[TUNI]{Jari J. J. Kangas}
\ead{jari.kangas@tuni.fi}
\ead[url]{http://www.tuni.fi}
\author[BU]{P. Robert Kotiuga}
\ead{prk@bu.edu}
\ead[url]{http://bu.edu}

%\cortext[corr]{Corresponding author, tel. +358-40-8490430.}
%\fntext[workdone]{}

\address[BU]{Department of Electrical and Computer Engineering, Boston University, Boston, MA, 02215 USA}
\address[TUNI]{Electrical Engineering, Tampere University, PO Box 1001, 33014 Tampere University, Finland}

\begin{abstract} \small
A wealth of literature exists on computing and visualizing cuts for the magnetic scalar potential of a current carrying conductor via Finite Element Methods (FEM) and harmonic maps
to the circle. By a cut we refer to an orientable surface bounded by a given current carrying path (such that the flux through it may be computed) that restricts contour integrals on a
curl-zero vector field to those that do not link the current-carrying path, analogous to branch cuts of complex analysis. This work is concerned with a study of a peculiar contour that illustrates topologically unintuitive aspects of cuts obtained from a trivial loop and raises questions about the notion of an optimal cut. Specifically, an unknotted curve that bounds
only high genus surfaces in its convex hull is analyzed. The current work considers the geometric realization as a current-carrying wire in order to construct a magnetic scalar
potential. Moreover, we consider the problem of choosing an energy functional on the space of maps, suggesting an algorithm for computing cuts via minimizing a conformally invariant functional utilizing Newton iteration.
\end{abstract}


\end{frontmatter}

% Our own commands
%\newcommand{\eqref}[1]{(\ref{#1})}
\newcommand{\fref}[1]{Fig.~\ref{#1}}
\newcommand{\rmd}{{\rm d}}
\newcommand{\TM}{{\mathrm{T}\Omega}}
\newcommand{\TxM}{{\mathrm{T}_x\Omega}}
\newcommand{\TstarM}{{\mathrm{T}^{*}\Omega}}

\newtheorem{mydef}{Definition}
\newtheorem{mytheorem}{Theorem}
\newtheorem{myremark}{Remark}

%\numberwithin{equation}{section}
%\numberwithin{myremark}{section}

%\linenumbers

\section{Introduction}

Benefits of obtaining topological information about a problem domain prior to solving electromagnetic 3D problems have been demonstrated and compared with standard methods in finite element modelling in numerous previous works. Here we consider particular surfaces, ``cuts,'' that offer information on the underlying domain's structure with the aim of improving a priori intuition about a solution's ``correctness.'' We consider that by using a ``good cut'' in some sense, we can improve the model and retain problem symmetries before transferring the topological information to more complex problems.

In our previous paper, we aimed to extend cuts for knotted geometries into undergraduate curricula via open-source software including Gmsh and Python in order to allow students to compute and 3D print surfaces \cite{CutsInEEEducation}. The exercises therein were intended to be a gateway to the intuitive study of near force-free magnetic fields and plasma physics \cite{KotiugaBlindTopologists}. Here we extend these methods to broaden the ability of students to utilize free, readily available tools to communicate technically through visualization. We consider a trivial loop as a motivating example of a ``good cut'' and offer an open-source approach to obtain solutions for arbitrary geometries of current-carrying (non-trivial) loops. We will then extend our example to a geometry homeomorphic to our initial trivial loop \cite{ThurstonAlmgren}, where we will see that even in this case the underlying problem geometry can lead to quite unintuitive results. We revisit the idea of a ``good cut'' and conclude with an extension of our methods for computing cuts utilizing a conformally invariant functional \cite{FreedmanHe}.

\subsection{Background and motivation}

The value of exposing a user to a cut utilized in obtaining a Finite Element solution to an electromagnetics problem is in itself a debate worth conducting; by its nature a philosophical debate that the authors have discussed  in earlier works \cite{CutsInEEEducation} and expanded upon here in Sections \ref{ValueOfCuts} and \ref{ValueToStudents}. In short, the authors have appealed in the past for employment of intelligible cuts as a useful tool in debugging software and assessing the structure of a problem domain. The authors have also posited that by posing the problem of obtaining a cut as equivalent to finding a solution of a magnetic scalar potential problem, one arrives at a relatively simple means to orient students in Electrical Engineering to the topological questions that underlay solutions returned by ``black box'' software packages. Temporarily foregoing further exposition on the matter, the authors operate assuming the following:
\begin{enumerate}
\item Cuts are useful. Obtaining additional information about how a calculation is performed and having opportunity to specifically indicate a user's preference in how the domain is split for calculation by FEM can only help and not hurt the quality of solutions. 
\item Intelligible cuts are more useful than non-intelligible cuts. One can distinguish between a cut that offers intuitive understanding of a problem domain and one that merely fits within the limits of a cut definition. \label{intelligible}
\end{enumerate} 
The focus of this exposition centralizes on the idea posited in Assumption \ref{intelligible}, above. The authors seek to point out a simple formulation of the cuts problem in terms of the magnetic scalar potential, appealing to a straightforward intuition that can and should be developed by those producing FEM software for problems in electromagnetics. The topological concepts undergirding the cut problem and the extension of these techniques in search of an ``optimal cut,'' should one exist, is less pedestrian and is not written for the non-initiated. The authors aim to frame the cut problem for those interested in engaging the topological structure discussed in their own research and/or with their students. In doing so, we attempt to develop a straightforward understanding of the cuts problem and nod towards the ``mass-appeal'' pedagogical advantages therein, while keeping our overarching focus centered on the specialist discussion offered by Assumption \ref{intelligible}. In particular, this exposition assumes familiarity with (co)homology groups and Lefschetz Duality for which we refer the reader to, e.g. \cite{PrkPwg}, for a practical review of the same in relation to electromagnetic phenomena. Where further specialist terminology arises in the text, the authors have cited useful reference material.

\subsection{The structure of the paper}

In Section~2, we discuss the notion of an optimal cut and review different aspects of optimality. In Section~3, we review the structure of the problem for finding cuts in terms of computational electromagnetics. We introduce our method for finding cuts and discuss its implementation. Section~4 demonstrates the method in action through computing a family of cuts for a trivial loop. In Section~5, we give another example and discuss the unintuitive surfaces arising from the Thurston-Almgren unknot. We propose to optimize in the space of maps against a conformally invariant functional and propose an algorithm for finding cuts utilizing such a functional and Newton iteration in Section~6. Finally, conclusions are drawn in Section~7.

\section{The quest for an optimal cut}

What is a ``good cut?'' Clearly, there is no simple answer to this rhetorical question. In this section, we briefly review some notions of optimality for cuts. Reflecting on the literature, the convoluted nature of the question is obvious.

\subsection{Optimal cuts: A world of trade-offs}

Here, we focus on four notions of optimality for cuts: 
\begin{enumerate}
\item Minimal genus.
\item Minimal surface area.
\item Minimal computational complexity.
\item Minimal energy.
\end{enumerate}

\noindent Optimizing the computation of cuts in terms of computational complexity has its obvious advantages, and minimal area cuts can lead to e.g. small supports for finite element basis functions utilizing such cuts, enhancing computational efficiency. The motivation for having genus minimizing cuts arises on one hand from knot theory and on the other hand it is related to near force-free magnetic fields where a connection to the Giroux correspondence yields minimal genus cuts, which have a direct interpretation in terms of magnetic scalar potentials \cite{KotiugaBlindTopologists}. Moreover, in order to have a ``nice'' cut, we might ask for certain further properties from the cuts. Are we fine with dealing in mere cycles or cocycles? Or should we rather opt for smooth manifolds, yielding us the tools of multivariable calculus? One can ask if there exists a notion of \emph{intuitively optimal cuts} and if it is related to any of these four notions of optimality; in some sense, this points to the direction of minimal energy cuts, as we shall see.

\subsubsection{Minimal area vs. minimal genus vs. computational complexity}

Computational efficiency of computing cuts for the needs of magnetoquasistatics (MQS) modelling has been studied for example in the works of Ruben Specogna (see e.g. \cite{Specogna2009, Specogna2008}). Efforts have been made for implementing general, automatic and efficient algorithms for such needs \cite{Dlotko}. Indeed, utilization of cuts combined with their efficient automatic computation can, for instance, be a very significant difference-maker in efficiency when solving non-linear MQS problems; for a recent example, see e.g. \cite{LahtinenJSNM}. However, the requirement of computational efficiency can be largely incompatible with intuitiviness of the resulting cut, let alone other notions of optimality. 

In general, there is no polynomial time algorithm to find minimal genus cuts. Moreover, Thurston and Almgren have showed that there is a trade-off between minimal area and minimal genus cuts \cite{ThurstonAlmgren}. More recently, Dunfield and Hirani studied the problem of minimal area cuts, and showed that in the special case of trivial second homology for a knot embedded in a 3-manifold, the least area surface bounded by the knot can be found in polynomial time \cite{DunfieldHirani}. As for finding minimal genus cuts, Agol, Hass and Thurston showed that the decision problem of whether a polygonal knot in a closed 3-manifold bounds a surface of genus at most $g_0$ is NP-complete \cite{AgolHassThurston}. Later on, they also showed that a similar decision problem, whether a curve bounds a surface of area less than some constant $c_0$ is NP-hard \citep{AgolHassThurston2}. For more studies of related problems, see works of Hass~\emph{et al.},  e.g. \cite{HassLagarias2004, HassLagarias1999, HassLagarias2003, HassThurston}. The problem of having the least number of triangles to span a polygon embedded in $\mathbb{R}^n$ is studied in \cite{HassLagarias2003}. It brings us to the context of finite element meshes, where we can also consider different mesh-dependent optimality criteria for cuts. For example, can we minimize the number of triangles in a cut? Seminal work of Wolfgang Haken on 3-manifold topology forms a starting point for such considerations \cite{Haken}. Indeed, it can be shown that getting within a certain percentage of the global minimum can be achieved in polynomial time.

In any case, it seems that different notions of optimality for cuts are incompatible. As shown in \cite{ThurstonAlmgren}, and as we will demonstrate via the working example in this paper, the following two criteria for calculated cuts are often not satisfied simultaneously, not even in a topologically trivial case:

\begin{enumerate}
\item Cuts are of minimal genus.
\item Cuts are of minimal surface area.
\end{enumerate}

\noindent As for Case~1, such a cut always exists, but there cannot be a general polynomial time algorithm to accomplish this. Case~2 seems to be possible in the simplest of cases, but there is no general algorithm for this. Moveover, as noted above, these conditions are often incompatible.

\subsubsection{Minimal energy}

Kotiuga's algorithm for computing cuts exploits the properties of the circle as an Eilenberg-MacLane space \cite{KotiugaCuts}, and the resulting co-dimension one cycles are Compact Orientable Embedded Manifolds with Boundary (COEMBs) \cite{KotiugaEMacCuts}. These are easily reconciled with the tools of multivariable calculus. Kotiuga goes further and advocates for putting an energy on the space of maps.

This brings us to the notion of minimal energy. Given the Dirichlet energy on the space of maps, finding the minimum of the Dirichlet energy
\begin{equation}\label{dirichlet}
\argmin_\phi E[\phi] = \argmin_\phi \frac{1}{2} \int_\Omega {|| \nabla \phi (x) ||}^2 \mathrm{d}V
\end{equation}
is equivalent to solving the Laplace problem for a map $\phi : \Omega \rightarrow S^1$ within a fixed homotopy class.
\begin{equation}\label{laplace}
\nabla^2 \phi = 0.
\end{equation}
In this context, cuts are level sets of scalar potentials solved from \eqref{dirichlet}. This fits hand-in-glove with the FEM and the physical interpretation in terms of magnetic scalar potentials. In this way, the computation of cuts reduces simply to FEM applied to the Dirichlet integral. However, as we will discuss later on in \ref{sec:EnergyFunctional}, the energy functional we will optimize against need not be \eqref{dirichlet}. A different choice of the energy functional may be utilized to calculate a different family of cuts, and we may select this functional according to properties of the cuts that we wish to emphasize. 

\subsection{Remarks on intuition and visualization} \label{ValueOfCuts}

In optimizing the cuts in terms of area, genus or computational efficiency, we have no guarantee that we will produce a visually intuitive cut. From the perspective of physical intuition, a family of cuts obtained by minimizing Dirichlet energy, guaranteed to be COEMBs, is useful as it is then e.g. easier to recognize how the ``same flux in magnetostatics'' traverses. Although the level sets of harmonic maps arise naturally in potential theory and are optimal in this sense, these cuts are still often difficult to interpret and considered ``unintuitive'' --- even by specialists in the field of computational electromagnetics. If we restrict ourselves to current-carrying knots and links, we can try make our cuts more intuitive by imposing additional geometric or topological conditions. These constraints are often trivial to impose in simple problems, but may be helpful in the context of more complicated situations.

A question of interest for us is: ``Should cuts be visually tangible?'' There are two answers that both seem obvious: yes and no. One might argue that a user of FEM software does not need to know about cuts, as long as she has the formulation she wishes at hand and arrives at a result. On the other hand, if the cut algorithm relates to physical or geometric intuition, a software developer might benefit from a visual representation e.g. for debugging purposes. Or, as is the case for magnetic fields and electromagnetics in general, if computing cuts is related to physics, physical insight can be gained from visualization.

\subsection{Optimality of cuts for learning and teaching} \label{ValueToStudents}

We have used the phrase ``Intuitively optimal cuts'' which is as
ambiguous as the phrase ``Optimality of cuts for learning and
teaching.'' Nevertheless, geometrical entities such as cuts are
inherent in the description of electromagnetic
phenomena. For example, inductance --- the quantity that relates flux
and current --- is often introduced in high school curricula. Cuts are
inherent in the definition of this basic parameter, which motivates the
question of how to best support learners progressing in an elementary EM
education.

As is well-known, there are differences in the ways students
learn. For example, Felder \cite{felder05:_under_studen_differ} lists
``three categories of diversity that have been shown to have important
implications for teaching and learning.'' Among these categories, differences in
students’ learning styles are predominant. For instance, as some students are `visual
learners' instructors should plan their teaching to ``find concrete
and visual ways to supplement the presentation of material that might
normally be presented entirely abstractly and verbally.'' Moreover,
some students are `active/concrete learners' who prefer ``engagement
in physical activity.'' Working with commonplace tools such a standard
FE software and 3D printing is one way to cater to these needs. In this case,
`printability' of cuts is of importance (i.e. whether cuts
are COEMBs).

Processes inherent in learning are discussed for example in Borovik's
book \cite{mathematics_under_microscope} which considers many aspects
of pedagogy, math, and cognition. In the following we have considered
only a few of them. Borovik lists two intertwined aspects of learning
and mastering mathematics:
\begin{enumerate}
\item The development of reproduction techniques for our own mental
  objects.
\item Interiorization of other people’s mental objects
  (i.e. visualization of abstract concepts, development of
  subconscious “parsing rules” for processing strings of symbols,
  etc.).
\end{enumerate}
Consequently, he raises concerns about the use of computer-assisted
learning---whether it dismisses ``cognitive content of standard
elementary routines'' that he considers as ``building blocks for
learning mathematics.'' Thus, in the context of this paper, we should
carefully decide how and when the commonplace tools are
used. Furthermore, use of the tools should strengthen and preferably
introduce new ``building blocks for learning electromagnetics''
instead of dismissing or hiding them.

Borovik also discusses the formation of mental images of real objects:
``When learning or doing mathematics, we quite frequently have to
create mental images of mathematical objects with eidetic qualities as
close to that of the images of real things as possible.'' This
suggests an obvious conclusion, i.e. that appropriate mental images
should be formed and that ``non-optimal cuts for learning and
teaching'' may hinder the building of proper understanding. Likewise,
intuition often relies strongly on first cases analyzed and ``wrongly
built intuition'' may hinder the building of understanding (see a
related discussion in \cite{KotiugaBlindTopologists} about the works
of visually-impaired topologists).

Equivalence classes are present even in elementary mathematics,
but challenges are often met when working with equivalence classes of
more complicated entities. Borovik states that, ``We can easily
visualize a collection of objects if we have seen them---one man or a
crowd, one flower or the whole garden in blossom. But it is very
difficult for a human to form a mental image of a multitude of
movements of his or her hand and treat this multitude a single
entity.'' This remark holds well also in the case of cuts and
inductance. Thus we propose that means to better visualize the multitude of
representatives of a class and the related equivalence relation likely
supports many learners. This possibility can be seen as one
benefit of defining cuts via levelset of scalar potential which allows
also the use of the commonplace tools in education.

\section{Review of problem structure}

For background information on the cuts problem, we refer the reader to \cite{PrkPwg}. We begin by considering a tubular neighborhood of some closed contour in $\mathbb{R}^3$. We take the complement of the conducting path within our problem region, and mesh over it to give a simplicial complex in $\mathbb{R}^3$. Interpreting this in terms of magnetostatics, the tubular neighbourhood of the contour is taken to be a current-carrying wire and we may set up a standard field problem to solve for the magnetic scalar potential. The procedure is standard: In the complement of our conducting region, The current density $J = \mathrm{curl}(H) = 0 \Rightarrow \mathrm{grad}(\phi) = H$, where $H$ is the magnetic field intensity and $\phi$ is the magnetic scalar potential. Observing closed contour integrals within the domain, the need for a cut is clear; a non-zero value of $\mathrm{curl}(H)$ results from any contour integral that links a non-zero current, and as such we can not have an associated single-valued potential function. We must partition the domain to effectively prevent current linkage by any contour integral within the mesh. 

Put simply, in our non-conducting region $\Omega \in \mathbb{R}^3$, we are asking topological information of the domain’s holes (the conductor). Specifically, we are trying to obtain a chain of 2-simplices whose boundary is the hole and/or the bounding sphere. In doing so, we are asking for a representative of the \emph{second relative homology group}, $H_2(\Omega, \partial \Omega)$. Technically, any representative generator of $H_2(\Omega, \partial \Omega)$ will do. Similarly, by Poincar\'e duality we could perform a cohomology computation to obtain a representative of the \emph{first cohomology group} $H^1(\Omega)$, and use it similarly. In terms of field computation, the choice of which group we choose to compute amounts only to a difference in how we propose to solve for our field. The former we identify as the so-called ``thin cut,'' and the latter the so called ``thick cut.''

Numerically speaking, in utilizing the thin cut, one must perform the homology computation on the domain indicated above using some --- ideally computationally effective --- algorithm. Once this computation is performed, one can double the nodes on the cut and reconnect the complex such that either side of the cut is not connected across the cut. One now has distinguished the two sides $S_{-}$ and $S_{+}$ of the surface $S$ representing the cut. Once the domain is split in this manner, one then introduces a potential jump 
\begin{equation}\label{jump}
\phi = \phi_1 \quad \mathrm{on} \quad S_{-}, \quad  \phi = \phi_2 \quad \mathrm{on} \quad S_{+}
\end{equation}
across the cut to account for a linkage around the conductor. Standard solvers can then be employed to solve for the scalar potential from \eqref{laplace}.

In utilizing thick cuts, one must similarly solve for $H^1(\Omega)$ using any standard algorithm. This returns representative cocycles (edge-based cohomology basis functions) that, in terms of computational electromagnetics, are then employed to account for currents flowing within the domain's holes (the conducting region). The cut is then utilized to solve for the scalar potential, e.g. as in \cite{LahtinenJSNM}.

In what follows, we employed a thin cut for use within our solver, however the same general framework applies for cohomology based computations.

\subsection{Computing the thin cut}

We employed strictly open-source software to implement all aspects of what follows and make all code freely available to those interested on GitHub\footnote{\url{https://github.com/AlexStockrahm/ThurstonAlmgren.git}}. We chose to write the geometry module in Python (\url{www.python.org}) with the aim of making these methods accessible to electrical engineering students with only basic knowledge of programming. 

\begin{figure}[!h]
\includegraphics[scale=.33] {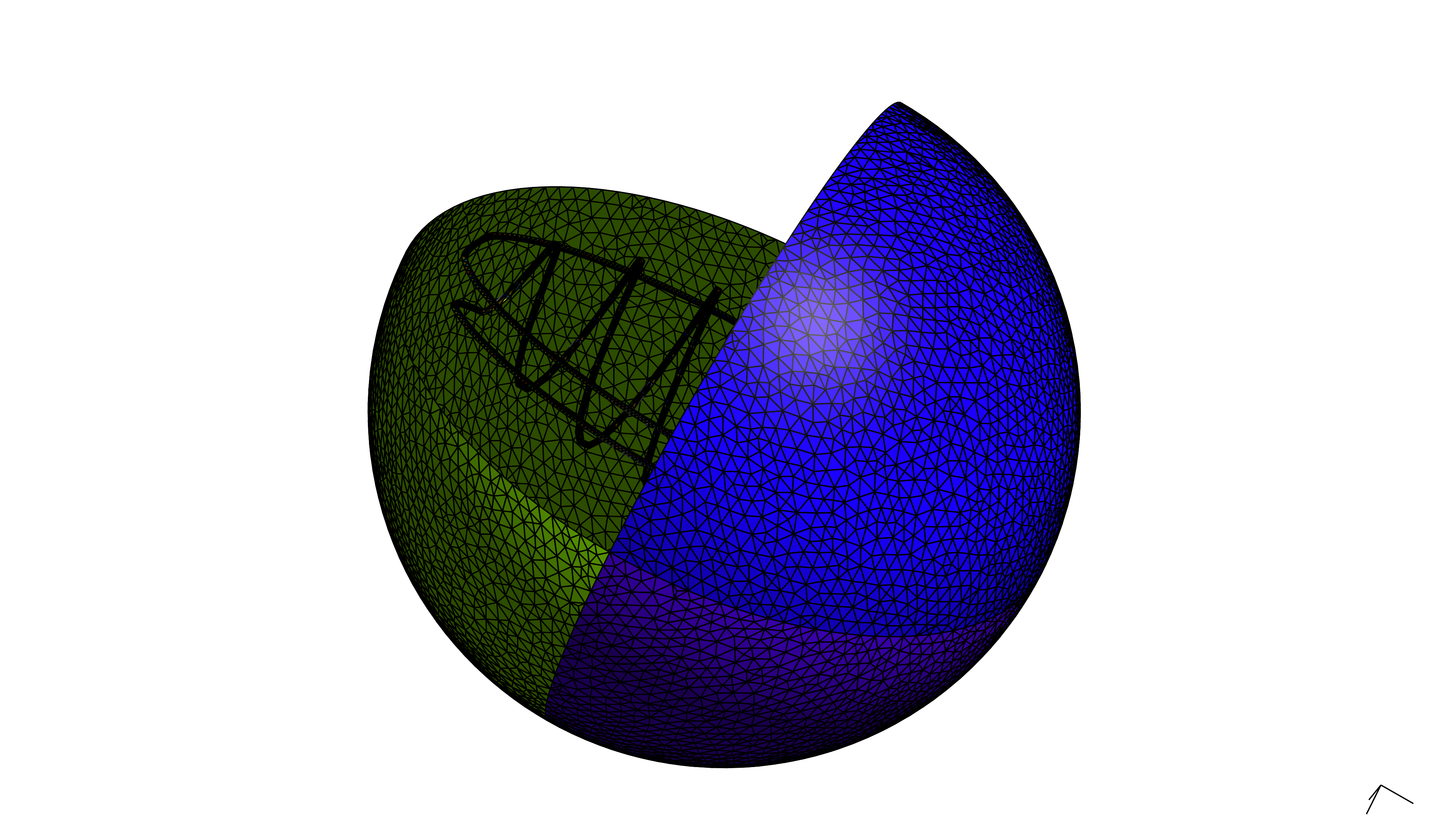}
\caption{A tubular neighborhood of a closed contour and the bounding sphere as parts of a finite element mesh.}
\label{fig:T-A-Geometry}
\end{figure}

A closed, one dimensional contour of any geometry in $\mathbb{R}^3$ is required as an input. This contour is then used to generate a tubular neighborhood of radius specified by the user, posing the problem with finite energy. A bounding sphere is then generated, and the geometry is written to an output Gmsh .geo file. Surface normals of the conducting path are taken inward while the bounding surface are taken outward such that the geometry characterizes the complement of the conducting region. An example of a contour and the bounding sphere is depicted in Figure~\ref{fig:T-A-Geometry}. 

The open-sourced finite element program Gmsh (\url{http://gmsh.info}) is then used to mesh the domain and perform the initial relative homology calculation. That is, we find a representative cycle of $(H_2(\Omega,\partial \Omega))$, using the on-board homology solver module of Gmsh and associated plugins.\footnote{Physical groups required by Gmsh are set during the geometry generation phase in Python such that system calls can be automated within Gmsh using a series of batch scripts, which we generated for computing cuts over many geometries, either in parallel or sequentially. It is also here that preferences for mesh density and related parameters can be set. Passing the physical group of the non-conducting volume ($\Omega$) to the homology solver and indicating the physical groups of the relative subdomains ($\partial \Omega$) i.e. the bounding sphere and the boundary of the conductor, generates a representative thin cut.} This cut is then duplicated using the \texttt{HomologyPostProcessing} plugin of Gmsh \cite{HomologyPlugin}. The resulting surfaces can then be used to split the domain as discussed above. Reconnection of the cut to either side of the domain is accomplished using the \texttt{Crack} plugin of Gmsh. The resulting mesh file, now containing all the necessary homological and domain-related information, may now be used to compute the scalar potential, by imposing a potential jump \eqref{jump} across the cut and solving \eqref{laplace} using FEM.

\section{Example: The trivial loop}

\begin{figure}[!h]
\includegraphics[scale=.33 ] {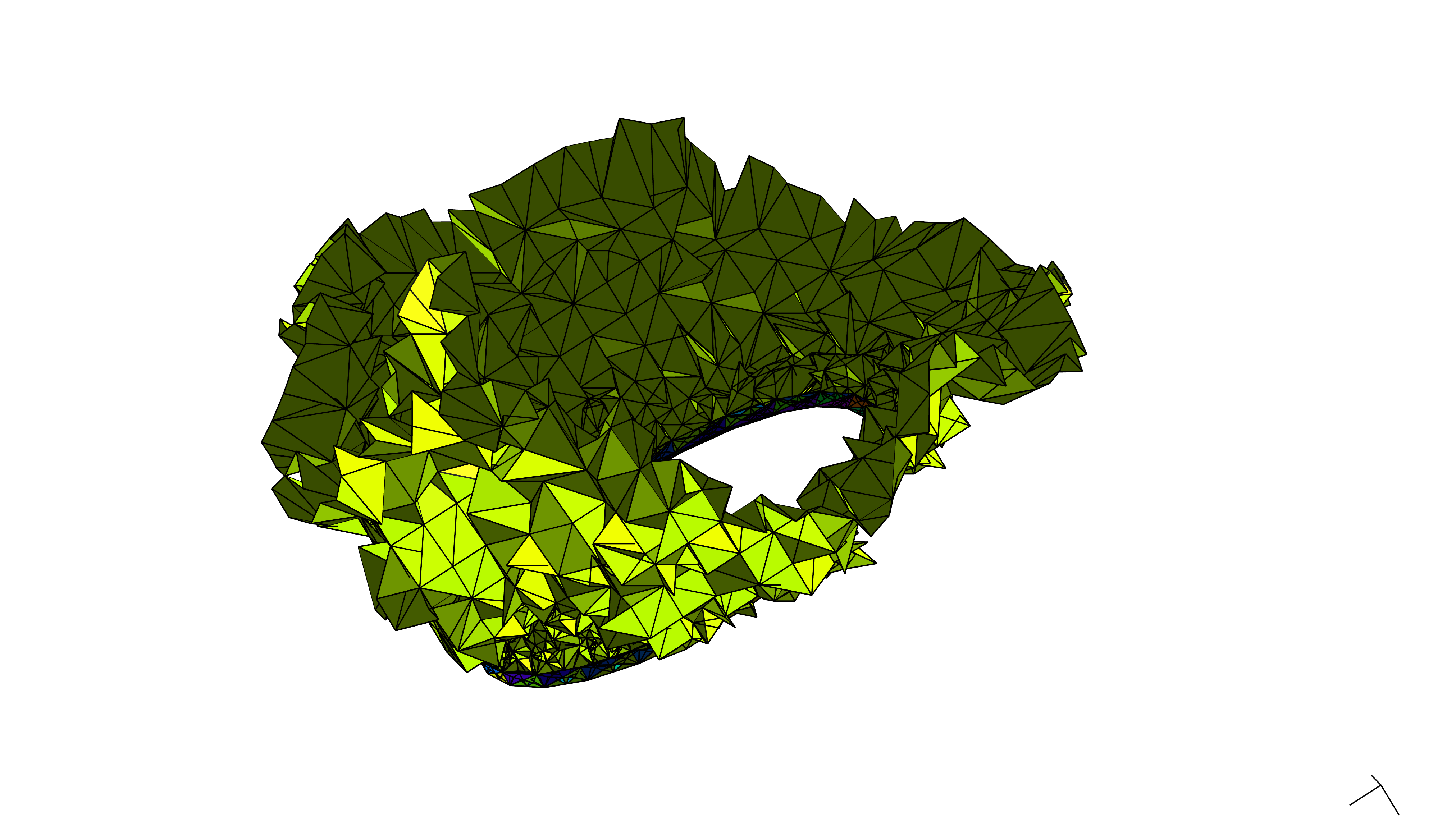}
\caption{Gmsh computed thin cut for a trivial loop. The computed 2-dimensional surface is bounded by an enclosing sphere and the trivial knot, the darker loop obscured by the sub-optimal cut but visible in the near and far side of the figure. The surface suitably partitions the domain for use in a magnetic scalar potential formulation, but fails to obtain agreeable characteristics.}
\label{fig:initialThin}
\end{figure}

We take the trivial loop as our first example. First, we generate the intial thin cut, depicted in Figure~\ref{fig:initialThin} using the integrated homology solver of Gmsh.

In reviewing this initial thin cut, we note that we have indeed obtained a member of $H_2(\Omega, \partial \Omega)$. However, we note the following shortcomings relative to what we might ask of a cut:
\begin{itemize}
\item The cut is not smooth
\item The cut is not minimum genus and not minimum area --- this view does not align with our intuitive notion of what this cut should obviously look like from an elementary electromagnetics education standpoint.
\item This cut does not preserve symmetries of the problem domain.
\end{itemize}

\noindent Simply stated, this cut appears unintuitive at best, and at worst it appears ``sloppy.'' We note also that as we are using the cut to establish a problem domain for a potential jump, and as such it plays an essential role in the quality of our solution; garbage into a model typically leads to garbage out. For this trivial case it makes sense to assess whether we can make a cut that is in some sense better, subject to any of the constraints above.

\begin{figure}[!ht]
  \centering \caption{Gmsh computed thick cut for a trivial loop on a coarse mesh, arrows indicating the computed representative first cohomology of the conducting region's complement.The left side of the computed cut encircles the conducting region, and traverses the loop's rim tracking the ``hole'' in the complement, i.e. the conducting region.}
  \subfloat[Top View]{\label{topView}\includegraphics[width=60mm]{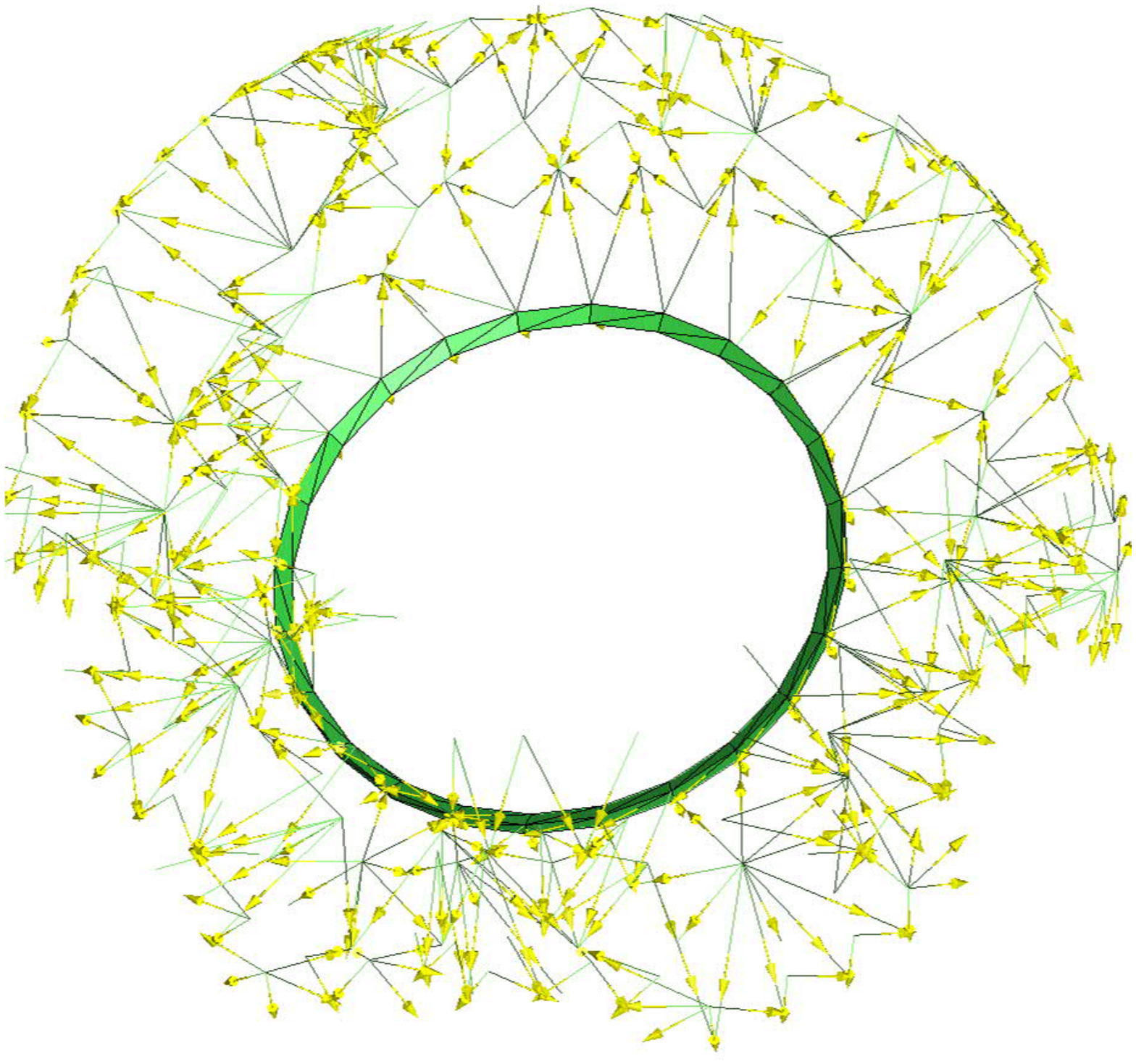}} \qquad
  \subfloat[Bottom View ]{\label{bottomView}\includegraphics[width=60mm]{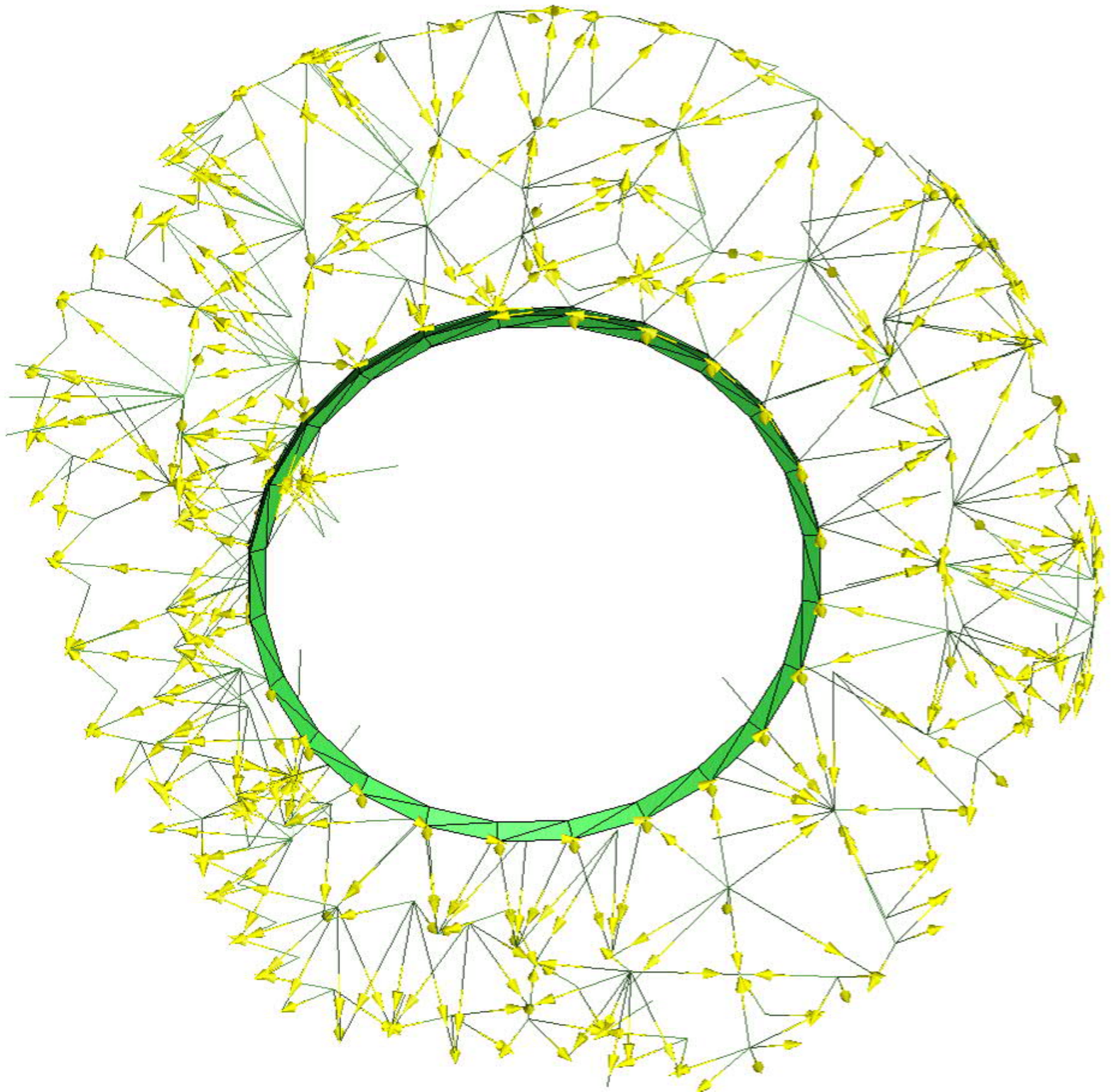}}
\label{thickCut}
\end{figure}

For demonstrative purposes we also calculate a thick cut for the case of the trivial loop, shown in Figure \ref{thickCut}. Accounting for the current flowing in the conducting region tracked by the cohomology representative allows a standard FEM formulation for finding the magnetic scalar potential with the possible benefit of not requiring a discontinuity imposition as in the case of the thin cut \cite{LahtinenJSNM}. However, the thick cuts are arguably less than intuitive in many cases. Being that the techniques discussed in this paper apply to improving the quality of \emph{thin} cuts, we sill set aside the use of the thick cut beyond acknowledging its duality to the thin cut and its utility in fixing the homotopy class of the set of maps to the circle, as noted in Section \ref{Functional}.

\subsection{Improving the thin cut}

We are able to efficiently obtain an orientable surface with boundary that can be used to solve a multitude of problems in e.g. computational electromagnetics. However, we may ask to improve our cut such that it:
\begin{enumerate}
\item Is Compact, Oriented, Embedded Manifold with Boundary (COEMB).
\item Is minimum genus or minimum area – where we also implicitly ask, can we have both?
\item Is smooth and simply interpreted for feasibility.
\item Is perhaps minimum energy, subject to some variational principle.
\end{enumerate}

\noindent In reviewing our wish-list above, we note that Item~4 is perhaps inclusive of the first three items for obvious reasons, and indicates an algorithm for obtaining such a so-called ``nice'' cut.

We propose taking the cut as a levelset of some harmonic map to the circle, representing the cut as the solution of an elliptic PDE. In the case of the image of the harmonic map being a circle, the ``angle'' of the map on the circle is a multivalued harmonic function over the domain (of the map). As such, it is amenable to standard finite element techniques once the value is fixed by an integer homology calculation which produces a cocycle specifying the homotopy class of the map.

The obvious advantage is that after the homotopy class is fixed, these problems are straightforward to solve and are understood as part of a standard FEM education in electrical engineering. We find that we can impose an energy on selected maps, and then optimize with respect to a norm of our choice. We also note that 1) and 3) above are going to be satisfied before beginning our calculation. In performing such a calculation, the entire family of cuts (levelsets) is computed, and as such we can gain information about our problem structure and perhaps select by inspection a cut that is in some way optimal.

As a first step, we take the natural choice here and optimize with respect to the Dirichlet Energy $E$. In doing so, we have changed our cut problem to a familiar problem, namely Laplace problem \eqref{laplace} for solving for scalar potential. 

However, from our prior exposition, we are well aware of the drawback – in order to find such a cut, we need an initial representative of $H_2(\Omega, \partial \Omega)$. Thus, to obtain (in this sense) a better cut, we must simply continue on our path of solving for the scalar potential. We can then use our improved cut, perhaps to create a better-structured domain on a more complex problem. That is, we need only solve a straightforward linear problem in statics once to obtain a quality family of cuts that are optimized in some sense. One benefit here is that we may then transfer our improved topological information to more difficult problems, such as nonlinear quasistatics problems where the problem cannot be reduced to solving an elliptic PDE via frequency-domain techniques.

\begin{figure}[!h] 
\includegraphics[scale=.33 ] {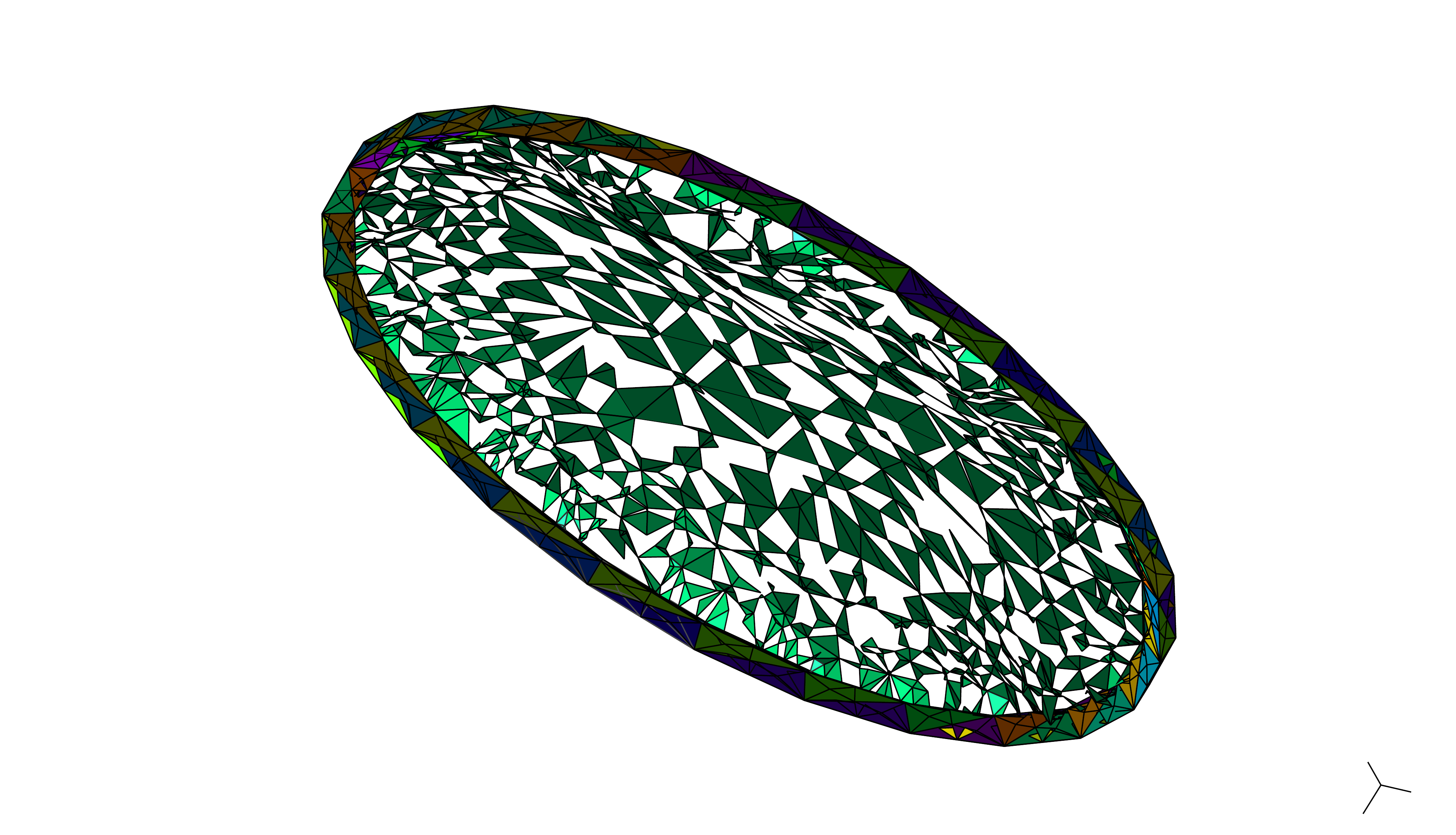} 
\caption{Thin cut for a trivial loop, obtained as a levelset of a scalar potential. Additional post-processing of levelset solutions was used in order to provide see-through portions of the cut for improved intelligibility when viewed on computer, crucial for more complex geometries (see Figure \ref{fig:T-A-Levelsets}). The difficulty in interpreting cuts in two dimensions is one of the main motivators for 3D printing the surfaces.}
\label{fig:thinCut}
\end{figure}

In order to obtain the levelsets of the scalar potential, we take our output mesh generated during the homology processing phase in Gmsh, and solve the Laplace problem in this domain using an external FEM based solver we wrote in Python. The results may then be returned to Gmsh or manipulated directly for post-processing and visualization of the levelsets. As expected, we generate a set of thin cuts on the trivial loop that hold our properties of interest. A member of the family is illustrated in Figure~\ref{fig:thinCut}.

We can see by inspection that we have satisfied the requests of obtaining a smooth COEMB, retaining problem symmetries and being minimal energy subject to our chosen energy functional. In short, we have a simple method for obtaining a cut that we might easily interpret as intuitive.

\section{The Thurston-Almgren Unknot and the Case of the Optimal Cut}

\begin{figure}[!ht]
  \centering\caption{The Thurston-Almgren Unknot}
  \subfloat[Seven helical twists are taken about the major axis, and a loop is taken around the twists, connecting back to the beginning of the contour.]{\label{figur:2}\includegraphics[width=120mm]{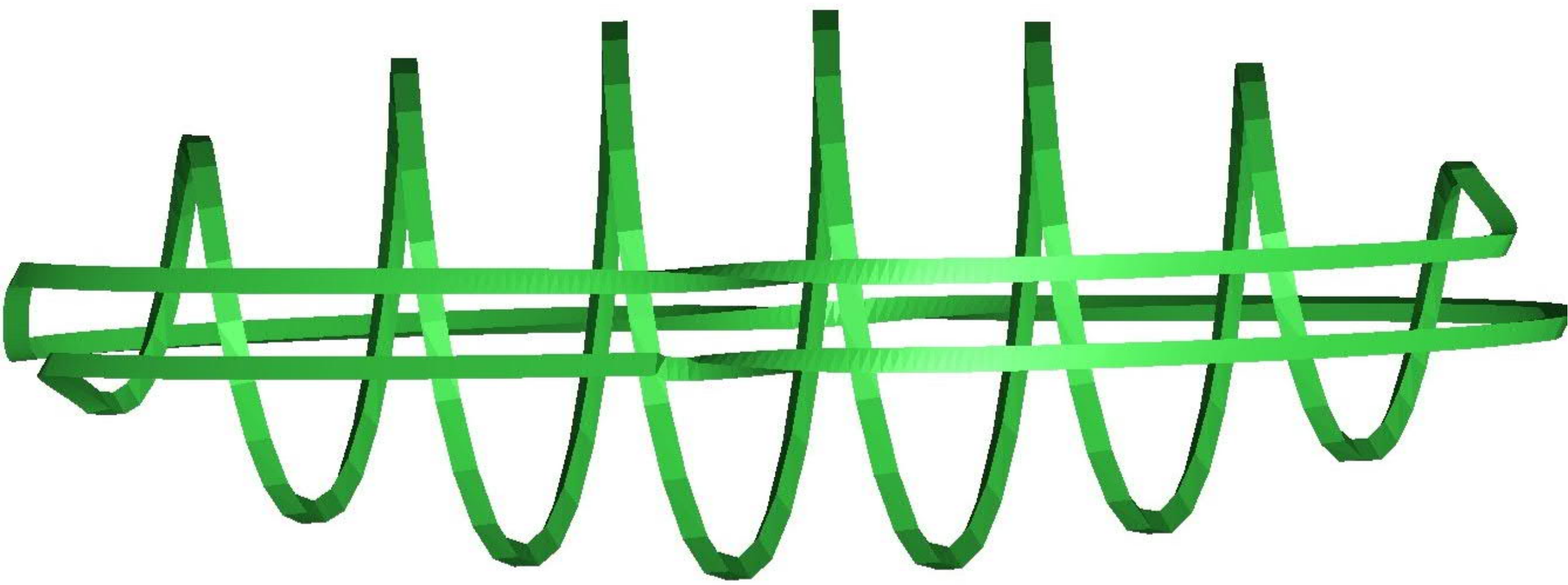}} 
\\
  \subfloat[A view of the contour with major axis vertical]{\label{figur:1}\includegraphics[width=60mm]{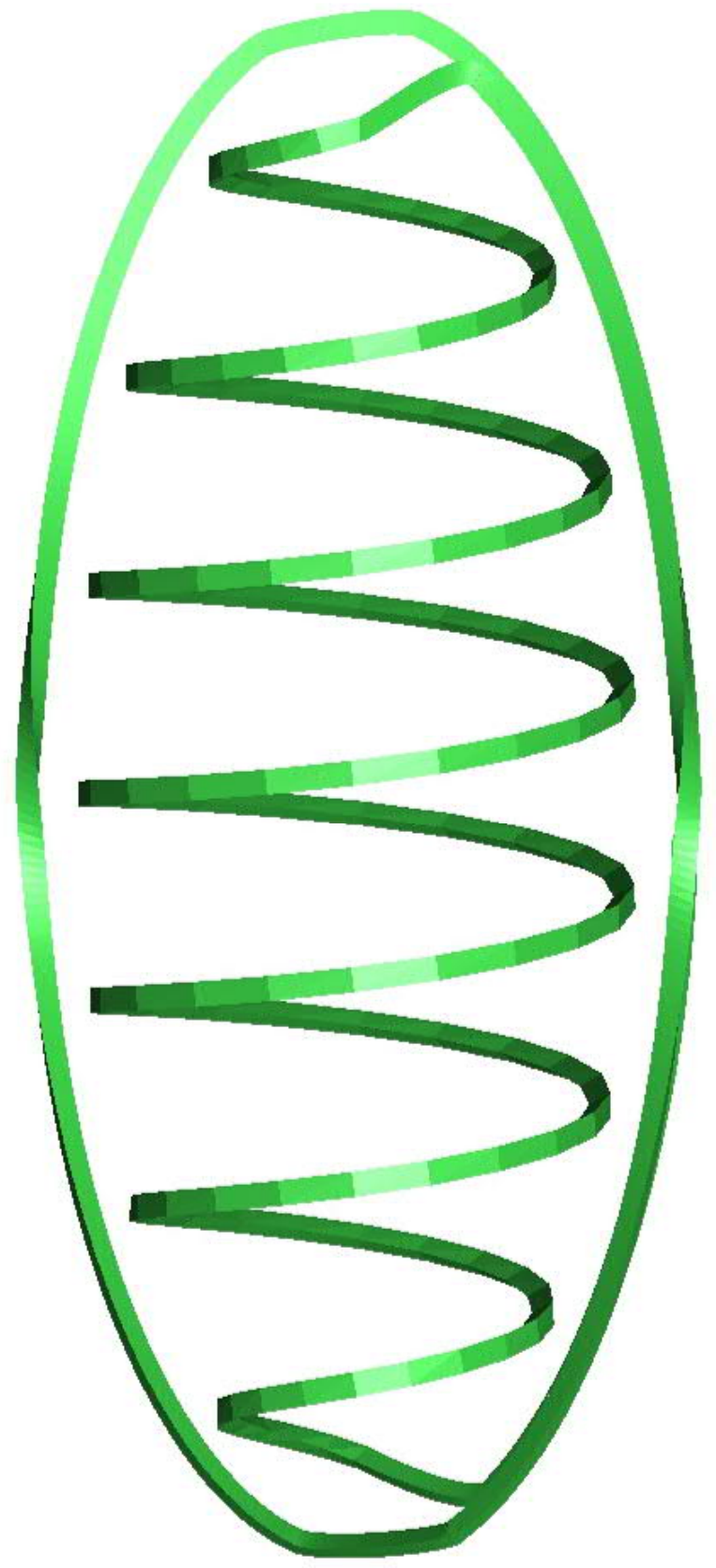}} \qquad
  \subfloat[The physics of the problem requires a wire diameter, $w$ and a minimum center-to-center distance, $d$. ]{\label{figur:3}\includegraphics[width=60mm]{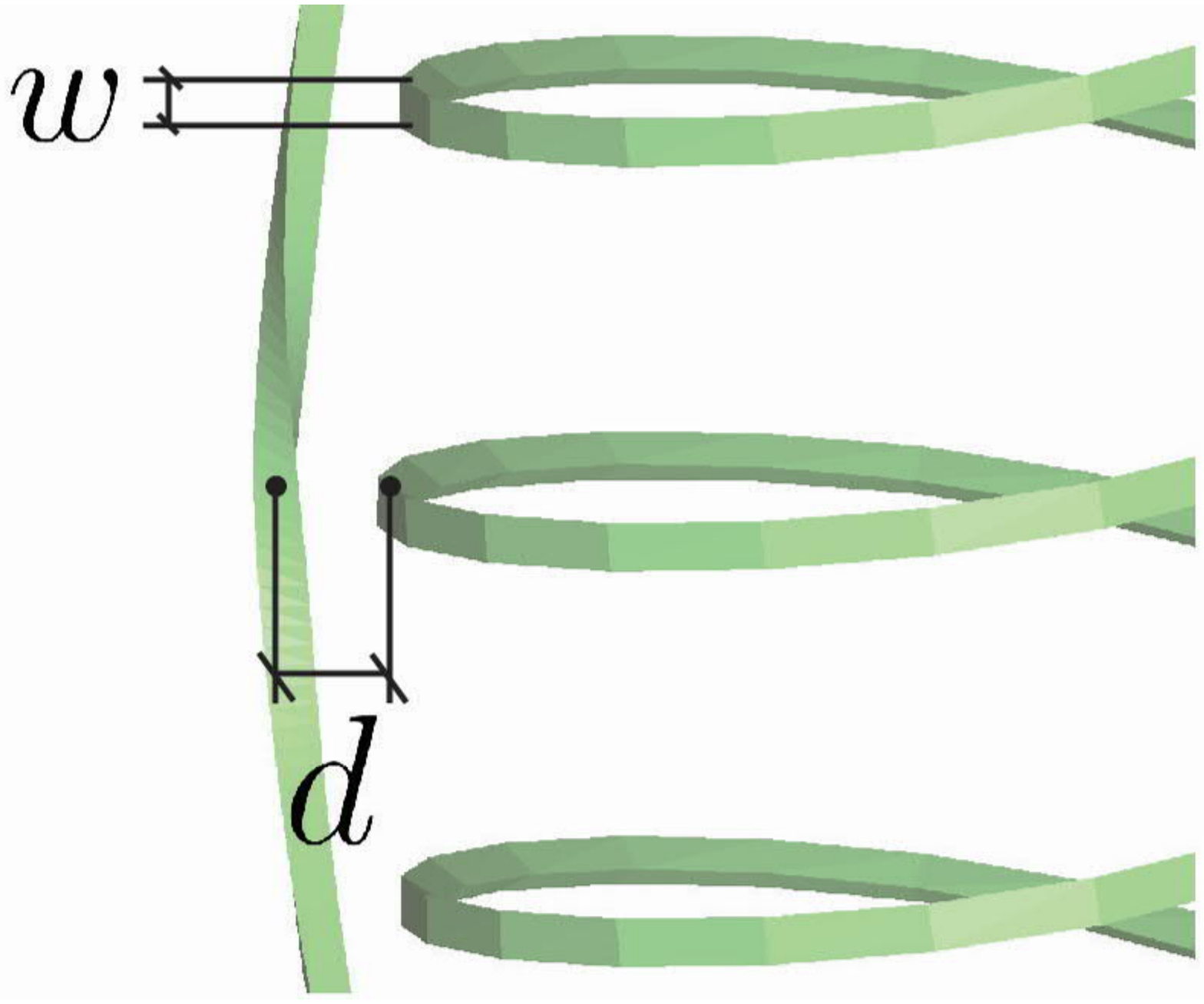}}
 \label{fig:T-A}
\end{figure}

We now inspect the case of the Thurston-Almgren Unknot: a topologically trivial loop that bounds only surfaces of high genus within its convex hull \cite{ThurstonAlmgren}. The curve is constructed by wrapping a helical contour around an ellipsoid in $\mathbb{R}^3$, and then revolving around the exterior of this contour along the ellipsoid's major axis, at first bypassing and then closing at the initial helix point, with a small separation maintained with the passing along the major axis. The number of helical twists around the ellipsoid characterizes the minimum genus of an oriented surface bounded within the convex hull. Specifically, for seven helical twists, the genus of any orientable surface bounded in the convex hull is at least 3, with increasing genus for additional twists. For the discussions that follow, we refer to specifically this 7-twist entity whose convex hull only bounds orientable surfaces with a genus of at least 3 --- whose characteristics are explored and shown in \cite{ThurstonAlmgren} --- which we chose to investigate. See Figure \ref{fig:T-A} for an example of the modeled Thurston-Almgren unknotted geometry.

As described by Thurston and Almgren, the geometry includes passings of the contour that are ``close'' to one another in a mathematical sense and it is only defined for a contour described as a 1-chain \cite{ThurstonAlmgren}. Hence, in order to model magnetic fields on this conductor geometry, it is necessary to give the wire some thickness and to mesh the interposing regions between passings (which can be made arbitrarily small).

Qualitatively speaking, this current path requires a great deal of approximation in terms of the FEM solution, as the field gradients tend to twist significantly, such that obtaining a sufficiently fine mesh is untenable. As a matter of practicality, the curve was parameterized in terms of the integer number helical twists, ellipsoid geometry, conducting region diameter, denoted $w$, and minimum passing distance, denoted $d$. For simplicity the parameter $\epsilon'=w/d$ was specified, where an $\epsilon'$ of unity implies touching conductors at crossing points, see Figure \ref{figur:3}. Bezier smoothing was used to interpolate between the different key turning points of the geometry. The characteristic length of the mesh, used within Gmsh to specify mesh granularity, was adapted based on $\epsilon'$. 

Using the tools developed, a series of 60 Thurston-Almgren geometries were generated and solved sequentially using a batch script over the course of 15 hours on a single node of the Boston University Shared Computing Cluster. We note that our code is in no way optimized as our main goal centralized only on proof-of-concept, and that our algorithms can be vastly improved for faster computation.

\begin{figure}[!h]
\includegraphics[scale=.33 ] {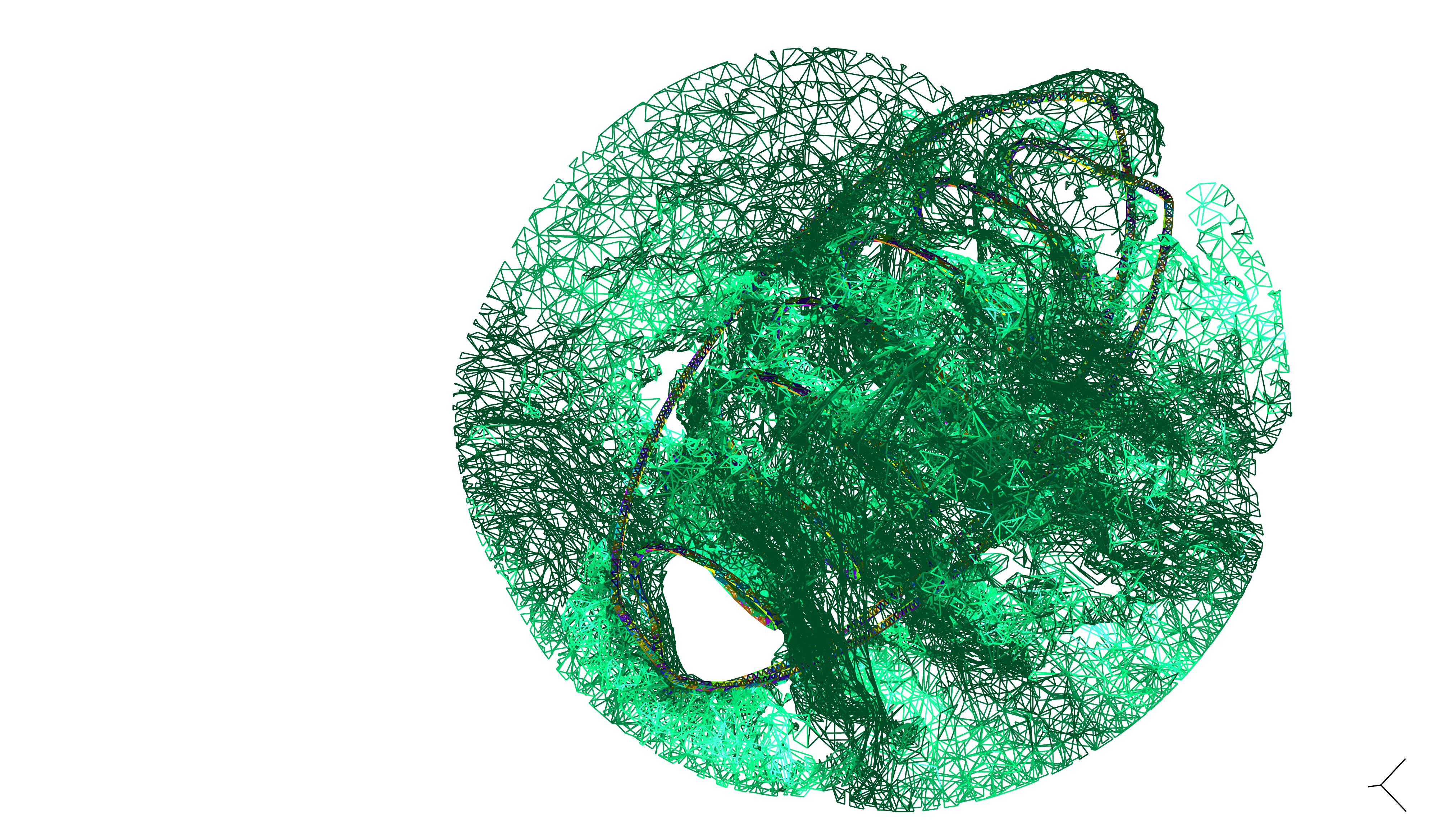}
\caption{Family of cuts for the Thurston-Almgren knot using the levelset method for the scalar potential.}
\label{fig:T-A-Levelsets}
\end{figure}

A family of cuts thus obtained is depicted in Figure~\ref{fig:T-A-Levelsets}. By the underlying property of bounding only high genus orientable surfaces in the convex hull, we are immediately led to our chief consideration of an ``optimal'' cut. Namely, a minimum area cut must live within the convex hull of this contour. If so, we are guaranteed to not have minimum genus (genus of such a surface for this geometry is at least 3) \cite{ThurstonAlmgren}. We see also that even though we can optimize our map of interest with respect to the Dirichlet energy and gain a smooth COEMB, the cut itself is still far from optimal in senses we might require. Indeed, understanding what an optimal cut might look like may require a new choice of energy functional, leading to a different solver. Further frustrating our efforts, we are only considering a geometry that is topologically equivalent (isotopic) to the initial trivial loop!

We see immediately from this case that even for ``simple'' geometries, finding optimal cuts has no single prescriptive method. Means of calculating the cut and selecting the best option is left to the decision-making of the modeler. It is however evident that, speaking generally, finding solutions of the cut problem as level-sets of elliptic PDEs offers a means of visually interpreting the underlying problem domain and understanding the underlying topology using a straightforward and relatively quick methodology. 

\section{On the choice of energy functional and norm}\label{Functional}

Use of the Dirichlet Energy was a natural choice as it conveniently formulates the cut problem in terms of the finite element method and a standard problem in statics. As we have shown, it is impossible to be sure that this cut method is the soundest option, even for trivial cases. We note here that in general, the choice of optimization affects only the implementation of a solver and that overall the methodology is unchanged.

In obtaining the initial thin cut we ask only for a representative that is equivalent homologically to any other cut that we can ask for, as homotopically equivalent spaces have isomorphic homology groups \cite{EilenbergSteenrod}. Similarly, in obtaining a thick cut we can construct any number of thin cuts using the methods presented above, acting on the dual space and using edge-based methods.

From our investigation of the Thurston-Almgren unknot, we have found that in selecting a particular functional we can only \emph{suggest} favorable properties for the resulting cuts. We find that even in topologically straightforward situations, we are left without any compelling reason to believe that a ``nice functional" implies a ``nice cut." However, once the homotopy class of the map is fixed, we are still able to formulate the cut problem in terms of any convex functional on the space of maps we might desire to investigate. This is to say, we may \emph{suggest} a functional's properties and can arrive at a means to solve for the cuts it yields via optimization. We provide a demonstration of this technique in the following subsection for a map with the property of conformal invariance.

\subsection{The conformally invariant functional and Newton's method} \label{sec:EnergyFunctional}

Let $I(u)$ a convex functional with bounded inverse Hessian. Then, Newton iteration for minimizing $I(u)$ is $u_{n+1} = u_n - \Delta_n$, where $\Delta_n$ is the solution of 
\begin{equation}\label{Delta_n}
\mathrm{Hess}(u_n;\Delta_n,v) = \delta I (u_n;v), \quad \forall v.
\end{equation}
In particular, if $I(u)$ is a functional consisting of $u$ and partial derivatives, then solving $\delta I(u,v) = 0$, equivalent to some Euler-Lagrange equations, and Newton iteration amounts to finding the minimum of the quadratic functional
\begin{equation}\label{eq:NewtonInf}
\frac{1}{2} \delta^2 I (u_n; \Delta_n, \Delta_n) + \delta I (u_n;\Delta_n)
\end{equation}
and setting $u_{n+1} = u_n - \Delta_n$.

Now, consider the following family of functionals:
\begin{equation}\label{gamaDirchlet}
\widetilde{I}_{\mu,\gamma}(\rm grad \phi) = \frac{1}{\gamma} \int_R \mu|\rm grad \phi|^{\gamma} dV,
\end{equation}
where $\gamma$ is our parameter of choice, $\mu$ is some positive weighting function, and $\phi$ is the potential function we seek. Here it is evident that setting $\gamma = 2$ corresponds to our Dirichlet integral with the inclusion of a weight term, which has an obvious interpretation in terms of a material parameter. The first variation of this functional is
\begin{equation}
\delta \widetilde{I}_{\mu,\gamma}(\rm grad \phi, \rm grad \delta\phi) = \int_R \mu|\rm grad \phi|^{(\gamma -2)}\langle \rm grad \phi , \rm grad \delta \phi \rangle dV.
\label{eq:FirstVariation}
\end{equation}

It is evident that if $\gamma$ is less than two, the resulting linearized Euler-Lagrange equation is not elliptic where $|\rm grad \phi|$ vanishes.

Computing the second variation of this functional we arrive at,

\begin{equation}
\begin{split}
\delta^{2} \widetilde{I}_{\mu, \gamma, 0}(\rm grad \phi, \rm grad \psi, \rm grad \psi) &\\
= \int_R &\mu|\rm grad \phi|^{(\gamma -4)}[|\rm grad \phi|^2\langle \rm grad \psi , \rm grad \psi \rangle \\
& + (\gamma-2)(\langle \rm grad \phi , \rm grad \psi \rangle)^2] dV
\label{eq:SecondVariation}
\end{split}
\end{equation}

In particular, a special case of \eqref{gamaDirchlet}, $\gamma=3$ yields the conformally invariant functional \cite{FreedmanHe},
\begin{equation}
I_{\mu, 3}(\rm grad \phi) = \frac{1}{3}\int_R \mu|\rm grad \phi|^{3} dV.
\label{gamaDirchlet3}
\end{equation}
Now, applying Newton iteration, we solve for
\begin{equation}
 \delta^2 I (u_n; \Delta_n, \delta(\Delta_n)) = - \delta I (u_n;\delta(\Delta_n)), 
\end{equation}
for all $\delta(\Delta_n)$, to find $\Delta_n$, and then set $u_{n+1} = u_n - \Delta_n$. For the conformally invariant functional \eqref{gamaDirchlet3}, equations \eqref{eq:NewtonInf}, \eqref{eq:FirstVariation} and \eqref{eq:SecondVariation} yield an explicit expression for the Newton iteration. This iteration can be used to either derive an Euler-Lagrange equation or a FEM iteration.

This sets the stage for implementing an algorithm for computing a new family of cuts, optimizing in the space of maps in terms of the conformally invariant functional \eqref{gamaDirchlet3}. This future work is expected to yield cuts which are more robust with respect to deformations of the mesh geometry. 

\section{Conclusions and outlook}

Cuts, special surfaces carrying topological information of the domain in question, are essential in computational electromagnetics and electrical engineering education. The notion of optimality of a cut is a multi-faceted issue and there is no simple answer to the rhetorical question, ``What is a good cut?'' In this paper, we reviewed different notions of optimality related to cuts and presented a workflow for obtaining cuts as levelsets of a scalar potential function. These cuts have many desirable properties, and are obtainable utilizing strictly open-source software. Furthermore, we demonstrated this workflow to illustrate unintuitive aspects of cuts for magnetic scalar potential arising from a topologically trivial loop considered as a current-carrying wire. Via this example, the incompatibility of the different notions of optimality was clearly exposed. Finally, we suggested finding cuts utilizing possibly nonlinear energy functionals that differ from the standard Dirichlet energy to further investigate the notion of an optimal cut. In particular, we suggested an algorithm for optimizing a conformally invariant functional utilizing a Newton iteration. This sets up the stage for implementing an algorithm for computing cuts, optimizing in the space of maps in terms of the conformally invariant functional as proposed. Moreover, the developed tools and examples presented provide a fertile soil for a pedagogical discussion.

\section*{Acknowledgment}

This research was partially supported by The Academy of Finland project [287027].

%\section*{References}
\section*{Declaration of Interest}

Declarations of interest: none.
\

\end{document}